\documentclass[%
reprint,
 amsmath,amssymb,
 aps,
]{revtex4-2}

\usepackage{graphicx}
\usepackage{dcolumn}
\usepackage{bm}

\begin{document}

\title{Detection of hidden photon dark matter using the direct excitation of transmon qubits}

\author{Shion Chen$^{(a)}$,
  Hajime Fukuda$^{(b)}$,
  Toshiaki Inada$^{(a)}$,
  Takeo Moroi$^{(b,c)\dagger}$,
  Tatsumi Nitta$^{(a)}$,
  Thanaporn Sichanugrist$^{(b)}$}

\affiliation{
$^{(a)}$International Center for Elementary Particle Physics (ICEPP), The University of Tokyo, 7-3-1 Hongo, Bunkyo-ku, Tokyo 113-0033, Japan
\\
$^{(b)}$Department of Physics, The University of Tokyo, 7-3-1 Hongo, Bunkyo-ku, Tokyo 113-0033, Japan
\\
$^{(c)}$QUP (WPI), KEK, Oho 1-1, Tsukuba, Ibaraki 305-0801, Japan
}

\begin{abstract}

  We propose a novel dark matter detection method utilizing the
  excitation of superconducting transmon qubits.  Assuming the hidden
  photon dark matter of a mass of $O(10)\ \mu{\rm eV}$, the classical
  wave-matter oscillation induces an effective ac electric field via
  the small kinetic mixing with the ordinary photon.  This serves as a
  coherent drive field for a qubit when it is resonant, evolving it
  from the ground state towards the first-excited state.  We evaluate
  the rate of such evolution and observable excitations in the
  measurements, as well as the search sensitivity to the hidden photon
  dark matter.  For a selected mass, 
  one can reach $\epsilon \sim 10^{-13}-10^{-12}$ 
  (where $\epsilon$ is the kinetic mixing parameter of the hidden photon) 
  with a few tens of seconds using a single standard transmon qubit.  
  A simple extension to the frequency-tunable SQUID-based transmon enables the mass scan to cover the range of $4-40\ \mu{\rm eV}$ ($1-10\ {\rm GHz}$)  within a reasonable length of run time.
  The scheme has great potential to extend the sensitivity towards various directions 
  including being incorporated into the cavity-based haloscope experiments or 
  the currently available multi-bit Noisy Intermediate-Scale Quantum (NISQ) computer machines.

\end{abstract}

\maketitle

\noindent
$^{\dagger}${\it Corresponding author.}

\vspace{1mm}
\noindent
\underline{\it Introduction}: After the discovery of the anomalous
velocity behavior of the galaxies in the Coma cluster reported by
F.~Zwicky~\cite{Zwicky:1933gu, Zwicky:1937zza}, 
the hypothetical mass source -- dark matter (DM) -- has been an outstanding mystery in physics.  
Even though various strong indirect evidences have been established through astrophysical and
cosmological observations, the absence of the direct detection
leaves its particle-physics properties still largely unknown.

Numerous efforts have been implemented for the direct DM detection.
Experiments using nuclear (electron) recoil processes provide
excellent sensitivities to DM of a mass around GeV to TeV (GeV to
sub-GeV) scale (for a dedicated review, see, e.g., 
Ref.~\cite{ParticleDataGroup:2022pth}). The recoil technique becomes less
effective for DM lighter than $\sim 10\ {\rm MeV}$ due to the recoil
energy below the detection threshold that is typically $O(1)\ {\rm keV}$. 

Different detection regimes are clearly needed for exploring lighter DM.
For DM weakly coupled to electromagnetic interaction (e.g., the hidden photon or axion),
haloscope experiments~\cite{sikivie1983experimental,sikivie2021invisible}  using microwave cavities
have provided the leading sensitivity below $O(1)\ {\rm meV}$~\cite{Nguyen:2019xuh,Godfrey:2021tvs,An:2022hhb,Cervantes:2022yzp,Dixit:2020ymh,agrawal2023stimulated,Brun:2019kak,Ramanathan:2022egk,DOSUE-RR:2022ise,ADMX:2001dbg,ADMX:2018gho,ADMX:2019uok,ADMX:2021nhd,ADMX:2018ogs,Choi:2020wyr,Jeong:2020cwz,Lee:2022mnc,Kim:2022hmg,CAPP:2020utb,HAYSTAC:2020kwv,Brubaker:2016ktl,TASEH:2022vvu,Alesini:2019ajt,Alesini:2020vny,Alesini:2022lnp,Quiskamp:2022pks,Cervantes:2022gtv,ADMX:2018ogs}.
Since such light DM can be treated as a classical matter-wave due to its high number density within their de Broglie wavelength,
the DM-converted photons can be accumulated in the matched resonant modes. 
This results in a detectably sizable electric signal read out from an antenna.

Alternatively, it has also been pointed out that the excitation processes in condensed-matter systems 
can be used to probe very light DM (see, e.g., Ref.~\cite{Mitridate:2022tnv}).  
Since the energy gaps in condensed-matter systems are generally
much smaller than $O(1)\ {\rm keV}$, the DM absorption 
may exhibit distinctive excitation signatures. 
The goal of this letter is to extend this idea to the superconducting quantum bits (qubits),
known as artificial atoms, by exploiting their favorable features: a strong coupling to electric fields, tunable energy gaps, 
and the precise readout/control functionality
enabling easy detection and manipulation of the excitation.
We focus on the two-level system of the lowest two energy states of the qubits, 
namely, the ground state ($|g\rangle$) and the first excited state ($|e\rangle$).
The energy gap is typically $O(1-10)\ \mu{\rm eV}$, corresponding to
the frequency of $O(1-10)\ {\rm GHz}$. The strong coupling allows 
the efficient DM absorption of a mass corresponding to the energy gap, 
driving the qubits from $|g\rangle$ to $|e\rangle$.  

In this study, we target the oscillating hidden photon field as the DM.
While being a well-motivated DM candidate,
it is also a natural constituent arising from a large class of string-inspired models in particle physics~\cite{Cicoli:2011yh}. 
Through the kinetic mixing with the photon, 
the hidden photon oscillation yields a weak coherent effective electromagnetic field which causes the $|g\rangle\leftrightarrow |e\rangle$ transition.
We estimate the transition rate from $|g\rangle$ to $|e\rangle$ driven by this feeble coherent field
assuming a standard transmon qubit architecture \cite{TransmonQubit}.
We find that, even with a single qubit in a typical specification of today's transmon, 
one can reasonably probe the unexplored parameter region of the hidden photon.
A wide frequency tunability ($\sim 1-10\ {\rm GHz}$) is easily acquired 
by extending to a transmon based on a Superconducting Quantum Interference Device (SQUID).
This is particularly remarkable given the much less required engineering effort 
compared to the typical haloscope experiments targeting the similar frequency range.
The sensitivity scalability along the number of qubits is also discussed for future prospects.

\vspace{1mm}
\noindent
\underline{\it Transmon qubit and hidden photon}: We start with introducing the interaction between a transmon qubit
and the hidden photon.  The natural unit (i.e., $c=\hbar=1$) is used throughout this letter.

A transmon qubit is modeled as a closed circuit loop consisting of a capacitor
element and a non-linear inductance realized by a Josephson junction or a SQUID.  
The Hamiltonian for the system is described by
\begin{align}
  {\cal H}_0 = 
  \frac{1}{2} C V^2 - J \cos \hat{\theta},
\end{align}
where $\hat{\theta}$ is the phase difference across the Josephson
junction, and $C$ is the capacitance. $J$ is positive-valued.
This is constant when a Josephson junction is considered as the inductance element, 
while it is tunable for a SQUID through the magnetic flux bias applied.
The voltage difference $V$ between the Josephson junction
is related to $\hat{\theta}$ as
\begin{align}
  V = (2e)^{-1} \dot{\hat{\theta}},
\end{align}
where $e$ is the electric charge of the electron.  The conjugate
momentum of $\hat{\theta}$, denoted as $\hat{n}$, is introduced for 
the canonical quantization, namely
\begin{align}
  \hat{n} \equiv Z\dot{\hat{\theta}},
\end{align}
with
\begin{align}
  Z \equiv (2e)^{-2} C.
\end{align}
Notice that $\hat{n}=CV/2e$ can be regarded as the total
charge in units of $2e$, and $\hat{\theta}$ and $\hat{n}$ satisfy the
commutation relation $[\hat{\theta}, \hat{n}]=i$.
The Hamiltonian is then written as
\begin{align}
  {\cal H}_0 = 
  \frac{1}{2Z} \hat{n}^2 - J \cos \hat{\theta}
  =
  \frac{1}{2C} (2e \hat{n})^2 - J \cos \hat{\theta}.
\end{align}
The energy levels of this system are unequally spaced; the ground and
the first excited states, denoted as $|g\rangle$ and $|e\rangle$,
respectively, are used for the transmon qubit.  The excitation energy
from $|g\rangle$ to $|e\rangle$ is denoted as $\omega$.  Then, the
Hamiltonian of the system is reduced to approximately
\begin{align}
  {\cal H}_0 = 
  \omega |e\rangle \langle e|.
\end{align}
For convenience of the later discussion, we also define
\begin{align}
  \hat{a} \equiv
  \frac{1}{\sqrt{2\omega Z}}
  \left(\hat{n} - i \omega Z \hat{\theta} \right),~~~
  \hat{a}^\dagger \equiv
  \frac{1}{\sqrt{2\omega Z}}
  \left( \hat{n} + i \omega Z \hat{\theta} \right),
\end{align}
which satisfies $[\hat{a},\hat{a}^\dagger]=1$.  
These correspond to the annihilation and creation operators
when approximating the potential as a parabolic one, 
where $J$ and $\omega$ are related as $J \simeq Z\omega^2$.

If there exists the hidden photon DM, an effective electric field is
induced by the hidden photon oscillation.  In the mass-eigenstate
basis, the interaction terms of the electron field $\Psi_e$ with
the electromagnetic (EM) photon $A_\mu$ and hidden photon
$X_\mu$ are given by
\begin{align}
  {\cal L}_{\rm int} = 
  e \bar{\Psi}_e \gamma^\mu
  \left( A_\mu + \epsilon X_\mu \right)
  \Psi_e,
\end{align}
where $\epsilon$ is the kinetic-mixing parameter.  Assuming that the DM consists only of the oscillating hidden photon, we denote the hidden photon field around the Earth as
\begin{align}
  \vec{X} = \bar{X} \vec{n}_X \cos m_X t,
  \label{Xosc}
\end{align}
where $m_X$ is the hidden photon mass, $\bar{X}$ is the amplitude of
the oscillation and $\vec{n}_X$ is the unit vector pointing to the
direction of $\vec{X}$.  The amplitude is related to the local
density of the DM as
\begin{align}
  \rho_{\rm DM} = \frac{1}{2} m_X^2 \bar{X}^2.
  \label{rhoDM}
\end{align}

The effective electric field which a qubit would sense is given by
\begin{align}
  \vec{E}^{\rm (eff)} = \vec{E}^{\rm (EM)} + \vec{E}^{(X)},
\end{align}
where $\vec{E}^{(X)}$ is the field induced by the DM,
and $\vec{E}^{\rm (EM)}$ is the reactive ordinary electric field induced by the metallic package surrounding the qubit that senses $\vec{E}^{(X)}$ (see the Appendix for the detailed discussion).
Using Eqs.\ \eqref{Xosc} and \eqref{rhoDM} one obtains:
\begin{align}
  \vec{E}^{(X)} = -\epsilon \dot{\vec{X}}
  = \bar{E}^{(X)} \vec{n}_X \sin m_X t,
  \label{Eeff}
\end{align}
with 
\begin{align}
  \bar{E}^{(X)} \equiv
  \epsilon m_X \bar{X}
  = \epsilon \sqrt{2 \rho_{\rm DM}}.
\end{align}

Hereafter, we consider the case that $\vec{E}^{\rm (EM)}$ and
$\vec{E}^{(X)}$ have the same time dependence; this is the case in
particular when considering a qubit located in a cavity-like metallic package.  
We parameterize the total effective electric field
as
\begin{align}
  \vec{E}^{\rm (eff)} = \bar{E}^{\rm (eff)} \vec{n}_E \sin m_X t,
\end{align}
where $\vec{n}_E$ is the unit vector pointing to the direction of
$\vec{E}^{\rm (eff)}$.
The ``package coefficient" $\kappa$ is introduced:
\begin{align}
    \kappa^{\rm}  \equiv \frac{\bar{E}^{\rm (eff)}}{\bar{E}^{(X)}},
    \label{kappa_def}
\end{align}
which is typically $O(1)$ or larger if the package size is $\gtrsim m_X^{-1}$ (see the Appendix for the details).

With the effective electric field, the voltage difference of the
capacitor becomes $V+d\bar{E}^{\rm (eff)}\cos\Theta\sin m_Xt$, with
$\Theta$ being the angle between $\vec{n}_E$ and the normal vector of
the conductor plate.  Concentrating on the terms up to the linear
order in $\epsilon$, the Hamiltonian is modified as ${\cal H} = {\cal
  H}_0 + \Delta {\cal H}$ such that
\begin{align}
  \Delta {\cal H} = C V d \bar{E}^{\rm (eff)} \cos\Theta \sin m_X t
  = 2 \eta  \sin m_X t (\hat{a} + \hat{a}^\dagger),
\end{align}
where $d$ is the effective distance between two conductor plates, and
\begin{align}
  \eta \equiv \frac{1}{2\sqrt{2}} d \bar{E}^{\rm (eff)} \sqrt{C \omega}
  \cos\Theta
  = 
  \frac{1}{2} \epsilon \kappa d \sqrt{C \omega \rho_{\rm DM}}
  \cos\Theta.
\end{align}

Calculating the matrix elements $\langle g|\Delta {\cal H}|e\rangle$ and $\langle e|\Delta
{\cal H}|g\rangle$ presuming that the excited state is well approximated as
$|e\rangle\simeq\hat{a}^\dagger|g\rangle$, 
the following effective Hamiltonian expression is 
obtained:
\begin{align}
  {\cal H} = 
  \omega |e\rangle \langle e|
  + 2 \eta \sin m_X t
  ( |e\rangle \langle g| + |g\rangle \langle e| ),
\end{align}
describing the interaction between the transmon qubit and the hidden photon DM.

\vspace{1mm}
\noindent
\underline{\it Time evolution of the qubit}:
Now we show that the DM-induced field $\bar{E}^{\rm (eff)}$ causes 
the Rabi oscillation of the qubit, a coherent drive between $|g\rangle$ and $|e\rangle$.
For a qubit state
\begin{align}
  |\Psi (t) \rangle = 
  \psi_g (t) |g\rangle + e^{-i\omega t} \psi_e (t) |e\rangle,
\end{align}
the time evolution is given by
\begin{align}
  i\frac{d}{dt} |\Psi (t) \rangle = 
  {\cal H} |\Psi (t) \rangle,
\end{align}
namely,
\begin{widetext}
\begin{align}
  i \frac{d}{dt}
  \left(
  \begin{array}{c}
    \psi_g \\ \psi_e
  \end{array}
  \right)
  =
  \left(
  \begin{array}{cc}
    0 & -i\eta (e^{-i(\omega-m_X) t}-e^{-i(\omega+m_X) t}) \\
    i\eta (e^{i(\omega-m_X) t}-e^{i(\omega+m_X) t}) & 0
  \end{array}
  \right)
  \left(
  \begin{array}{c}
    \psi_g \\ \psi_e
  \end{array}
  \right).
\end{align}
\end{widetext}

Suppose that the qubit frequency is tuned to be equal to the hidden
photon mass, i.e., $\omega=m_X$.  Neglecting the fast oscillating
component (rotation wave approximation), the evolution equation
reduces to
\begin{align}
  i \frac{d}{dt}
  \left(
  \begin{array}{c}
    \psi_g \\ \psi_e
  \end{array}
  \right)
  \simeq
  \left(
  \begin{array}{cc}
    0 & -i \eta \\ i \eta & 0
  \end{array}
  \right)
  \left(
  \begin{array}{c}
    \psi_g \\ \psi_e
  \end{array}
  \right).
\end{align}
Assuming that the qubit is initially at the ground state, i.e.,
$\psi_g(0)=1$ and $\psi_e(0)=0$, we obtain
\begin{align}
  \psi_g (t) \simeq \cos \eta t,~~~
  \psi_e (t) \simeq \sin \eta t.
\end{align}
The transition probability from the ground state to the excited
state is 
$p_{ge} (t)=|\psi_e (t)|^2\simeq \sin^2 \eta t$,
corresponding to a Rabi oscillation at a frequency of $\eta$.

Note that the discussion above is valid only within
the coherence time of the system $\tau$, i.e., $t < \tau$.
The coherence time can be defined for 
the DM and the qubit individually ($\tau_X$ and $\tau_{\rm q}$, respectively).
The former is estimated to be $\tau_X\sim
2\pi/m_X v_X^2$ with $v_X\sim 10^{-3}$ being the hidden photon velocity.
For the latter, the longitudinal coherence time $T_1$ is relevant here
since the dephasing is highly suppressed in the transmon limit ($JZ\gg 1$)~\cite{TransmonQubit}. 
A $T_1$ of $\sim 100~\mu {\rm s}$ is commonly achieved in the recent
experiments~\cite{kjaergaard2020superconducting,place2021new,wang2022towards}.
The coherence time for the system is dictated by the shorter one,
i.e., $\tau \simeq \mbox{min} (\tau_X, \tau_{\rm q})$, which is usually 
given by that of the qubit. Hereafter, we parameterize $\tau\equiv2\pi Q/\omega$, with
$Q$ being the quality factor.

Assuming that $\tau\ll\eta^{-1}$, the transition
probability from $|g\rangle$ to $|e\rangle$ within the coherence time
is evaluated as
\begin{align}
  p_*\equiv p_{ge} (\tau) \simeq (\eta \tau)^2.
\end{align}
Numerically, the transition probability is given by
\begin{align}
  p_{ge} (\tau) \simeq &\, 0.12 \times
  \kappa^2 \cos^2 \Theta
  \left( \frac{\epsilon}{10^{-11}} \right)^2
  \left( \frac{f}{1\ {\rm GHz}} \right)
  \nonumber \\ &\, \times
  \left( \frac{\tau}{100\ \mu{\rm s}} \right)^2
  \left( \frac{C}{0.1\ {\rm pF}} \right)  
   \left( \frac{d}{100\ \mu{\rm m}} \right)^2
   \nonumber \\ &\, \times
   \left( \frac{\rho_{\rm DM}}{0.45\ {\rm GeV/cm}^3} \right),
\label{p_ge}
\end{align}
where $f\equiv\omega/2\pi$, which is
related to the hidden photon mass as
\begin{align}
  f \simeq 0.24\ {\rm GHz} \times
  \left( \frac{m_X}{1\ \mu{\rm eV}} \right)
\end{align}
when $\omega=m_X$.

Note that the scheme is agnostic to the concept of ``detection volume'' since 
the excitation rate solely depends on the local electric field around the qubits.
At the high-frequency regime this is a distinct advantage over the cavity-based haloscope experiments 
where the signal power is suppressed by $\sim 1/f^3$ due to the dwindling resonant radius.

Two methods are possible for detecting the excitation; for searching large $\epsilon$ yielding substantially high excitation rate ($p_* > O(10\%)$) the characteristic Rabi modulation can be measured that can be easily differentiated from the noise; for smaller $\epsilon$, the repetitive counting experiments (further discussed in the following section) is powerful despite the higher dark counts, allowing one to probe $p_*$ as low as $10^{-4}$.

\vspace{1mm}
\noindent
\underline{\it Experimental setup and the measurement protocol}: 
A typical setup for transmon measurements~\cite{gao2021practical} is adopted in the search.
The transmon is assumed to be in the X-mon~\cite{barends2013coherent} or the conventional dumbbell~\cite{deng2023titanium} architecture,
which is either packaged with a Coplanar Waveguide (CPW) resonator on a chip surrounded by a metallic shield 
or housed in a microwave cavity~\cite{Paik3DImplementation} for the readout. 
The readout resonator is fabricated so that the detuning from the qubit frequency is within a few GHz,
for which multiple samples can be created to cover a wide frequency scan range.
The sample is thermally attached to the mixing chamber plate of the dilution refrigerator and cooled down to $\sim 10$~mK.
The dispersive readout~\cite{wallraff2005approaching} is used to determine the qubit state.
A sequential amplification by the quantum amplifiers (e.g., JPA~\cite{aumentado2020superconducting} or JTWPA~\cite{macklin2015near}) at 10~mK stage, low-temperature amplifiers (e.g. HEMT~\cite{HEMT}) at the 4K stage, and the ordinary room temperature amplifiers is used to ensure a good single-shot readout fidelity.

A repetitive counting experiment is employed to detect small excitation rate $p^*$ using the protocol as follow,
using a single or multiple SQUID-based transmon qubit(s) tuned to the same frequency.
For the multiple qubit setup, the independent readout are considered,
mainly assuming the operation in the currently available frequency-tunable NISQ machines (e.g., Google Sycamore~\cite{google_roadmap}).

\begin{itemize}
\item[1.] All the qubits are set to the ground state at $t=0$,
  and are left for evolution until $t=\tau$.  
\item[2.] A single-shot state readout for all the qubits. A $O(100)~{\rm ns}$ of readout time is assumed.
\item[3.] For a given $\omega$, the step 1--2 are repeated $n_{\rm
  rep}$ times. The total number of the qubit measurements is
    $N_{\rm try} \equiv n_{\rm q} n_{\rm rep}$ where $n_{\rm q}$ is the qubit multiplicity.  
   The number of expected excitations is $N_{\rm sig} = p_* N_{\rm try}$,
   ignoring the inter-qubit interference.
\item[4.] Repeat the step 3 in different qubit frequencies. 
\end{itemize}
The search is a resonant-finding where the DM signal exhibits 
a sharp peak on the continuum baseline in the excitation rate,
at a particular qubit frequency corresponding to the DM mass.

The false positive excitations (``dark counts'') are critical in this counting experiment regime.
The dominant source of dark counts is expected to be the qubit readout error 
given that the state-of-art single-shot ground-state readout fidelity is about 99.9\ \% 
utilizing the higher excited states~\cite{Chen:2022frn}.
A flat 0.1\ \% readout error is therefore assumed in later discussion,
however note that the figure can be still improved in a more optimal setup (e.g. by introducing quantum amplifiers,
or in the future driven by the high demand of improvement from the field of superconducting quantum computers).

Thermal excitation can be also a significant source of dark counts particularly at the lower frequency.
The expected number of such dark counts is
\begin{align}
  N_{\rm dark} = e^{-\omega/T} N_{\rm try}.
\end{align}
with $T$ being the temperature of the qubits.
In a typical dilution refrigerator, $T=10$~mK can be solidly achieved.  
One may even expect $T \sim 1$~mK or lower by introducing Continuous
Nuclear Demagnetization Refrigerator (CNDR)~\cite{Toda_2018}.  
However, it has been argued that the actual temperature of the qubits may be different due to
the quasi-particle creation processes on the qubit chip
where an excitation equivalent to a thermal residual of $\sim
30~{\rm mK}$ is reported~\cite{Jin_QubitTemperature}.  Hereafter, we assume two
temperature scenarios, i.e., 30~mK as the conservative estimate based
on the currently available setup, and 1~mK assuming the usage of CNDR
and the circumvention of the quasi-particle problem in the future.

It is also worth pointing out that the level of dark counts can be measured and validated using the ``sideband" 
given the nature of resonant-finding searches
where signals can only peak at specific frequencies.

\vspace{1mm}
\noindent 
\underline{\it Sensitivity estimation}:
The signal rate is calculated from Eq.\ \eqref{p_ge}
with $C=0.1\ {\rm pF}$, $d=100\ \mu{\rm m}$,
$Q=10^6$, 
$\kappa=1$,
$\rho_{\rm DM}=0.45\ {\rm GeV}/{\rm cm}^3$,
and with the angular average $\cos^2\Theta\rightarrow\frac{1}{3}$ being taken.
The qubit multiplicity is assumed to be either $n_{\rm q}=1$ or $100$ ignoring the inter-qubit interaction.  
For the dark counts, a flat 0.1\% readout error and the thermal excitation ($T=30\ {\rm mK}$ or $1\ {\rm mK}$) 
are considered.
The search sensitivity is derived by comparing $N_{\rm sig}$ and $N_{\rm dark}$.
Here, we apply a simple form of $N_{\rm sig}/\sqrt{N_{\rm dark}}$ 
as the proxy to the significance in the unit of Gaussian-equivalent standard deviation.
The following criterion is used for the DM detection in the study:
\begin{align}
  N_{\rm sig} > \mbox{max} (3, 5 \sqrt{N_{\rm dark}}),
  \label{signalobs}
\end{align}
which requires either $5\sigma$ in case of substantial amount of dark counts,
or minimum 3 signal events in a highly dark-count-free regime.

The qubit frequency scan is considered in the range of $1\leq f\leq
10\ {\rm GHz}$, corresponding to the DM mass of $4-40\ \mu {\rm eV}$.
A constant $Q=10^6$ is assumed across the frequency range, where the
scan step width $\Delta\omega=\omega/Q$ and the coherence time $\tau =
2\pi Q/ \omega$ are variable; the coherence time of the qubit is
assumed to be $O(100)\ {\rm \mu s}$.  The number of the scan points is
$\sim Q \log{10} \sim 2\times 10^6$.  We consider one-year operation
for the scan, then the time available for each scan point is $\sim 14$
seconds.  The readout time ($O(100~{\rm ns})$) and the interval between
the readout ($10~\mu {\rm s}$
\cite{Dixit:2020ymh}) are neglected as they are short enough compared
with the coherence time $\tau$.
\begin{figure}[t]
  \centering
  \includegraphics[width=0.95\linewidth]{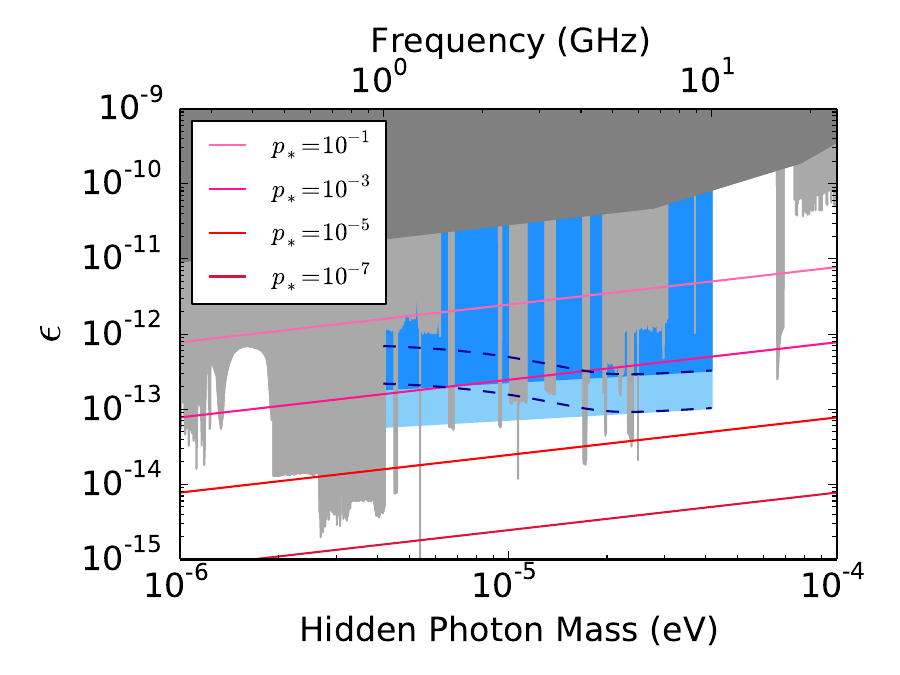}
  \caption{Contours of constant $p_*\equiv p_{ge}(\tau)$ on $m_X$
    vs.\ $\epsilon$ plane ($10^{-1}$, $10^{-3}$, $10^{-5}$, and
    $10^{-7}$, from the top). Parameters of $C=0.1\ {\rm
     pF}$, $d=100\ \mu{\rm m}$, $Q=10^6$, $\kappa=1$
    and $\rho_{\rm DM}=0.45\ {\rm GeV}/{\rm cm}^3$ are assumed.
    The gray-shaded region is excluded by
    the cosmological and astrophysical constraints
    \cite{McDermott:2019lch} (dark gray) and the existing hidden-photon search experiments \cite{Nguyen:2019xuh,Godfrey:2021tvs,An:2022hhb,Cervantes:2022yzp,Dixit:2020ymh,Brun:2019kak,Ramanathan:2022egk,DOSUE-RR:2022ise,ADMX:2001dbg,ADMX:2018gho,ADMX:2019uok,ADMX:2021nhd,ADMX:2018ogs,Choi:2020wyr,Jeong:2020cwz,Lee:2022mnc,Kim:2022hmg,CAPP:2020utb,HAYSTAC:2020kwv,Brubaker:2016ktl,TASEH:2022vvu,Alesini:2019ajt,Alesini:2020vny,Alesini:2022lnp,Quiskamp:2022pks,Cervantes:2022gtv,ADMX:2018ogs} (light gray) based on the summary in Ref.~\cite{Caputo:2021eaa}.  
    The blue-shaded regions indicate the
    sensitivity with the 1-year scan over the frequency range for
    $n_{\rm q}=1$ (dark blue) and $100$ (light blue) (more details in the main
    text) assuming the thermal noise of $T=1\ {\rm mK}$.
    The dashed lines show the
    sensitivity with $T=30\ {\rm mK}$ with the top (bottom) 
    line corresponding to $n_{\rm q}=1$ ($100$) respectively.
    A flat readout error of $0.1\ \%$ is assumed.
  }
  \label{fig:epsbound}
\end{figure}
Fig.\ \ref{fig:epsbound} shows the projected sensitivity of our
proposed experiment.

\vspace{1mm}
\noindent
\underline{\it Conclusions and discussion}: In this letter, we have proposed
a new detection scheme for the hidden photon DM 
using transmon qubits.  
Due to the small kinetic mixing with the ordinary EM photon, an
effective ac electric field is induced that coherently drives a transmon qubit
from the ground state toward the first-excited state when it is resonant.
We have calculated the rate of such excitation  (see Eq.\ \eqref{p_ge}), and evaluated the 
hidden DM search sensitivity. 
Using a standard SQUID-based transmon,
the sensitivity can reach $\epsilon \sim 10^{-13}-10^{-12}$ 
with $\sim 14$ seconds spent for a single frequency,
and with one year to complete the scan over the $4-40\ \mu{\rm eV}$ 
($1-10$\ GHz)
range.

While the depth of the sensitivity is generally weaker than the haloscope experiments using the cavity-resonance,
the proposed method is advantageous for the easier frequency tunability.
This is important feature for a fast ``shallow search'' targeting $\epsilon \sim 10^{-13}-10^{-12}$,
which is motivated by the fact that most of the cosmologically allowed frequencies have not yet been probed by direct search experiments.
The proposed method also has no less sensitivity compared with the other wide-band searches
using horn antennas~\cite{DOSUE-RR:2022ise}
or proposals utilizing the condensed-matter excitations (e.g., electric excitations
\cite{Hochberg:2016ajh, Hochberg:2016sqx, Hochberg:2017wce,
  Arvanitaki:2017nhi, Mitridate:2021ctr}, phonon \cite{Knapen:2017ekk,
  Griffin:2018bjn}, magnon \cite{Chigusa:2020gfs}, and
condensed-matter axion \cite{Marsh:2018dlj, Schutte-Engel:2021bqm,
  Chigusa:2021mci}).

There are a few considerations left for future studies that can further extend the sensitivity.
(1) Qubit design optimization maximizing the electric dipole moment,
where more aggressive transmon parameters and complex circuit design can be sought.
(2) Extend the frequency range beyond that typically explored by superconducting qubit experiments. 
0.2--20~GHz can be achieved with more dedicated RF setups.
(3) The $\kappa$ enhancement by the cavity resonance discussed in Appendix can be further investigated, particularly in the context of being incorporated into the haloscope experiments.
The setup would be similar to the experiment performed by A.~Dixit et al.~\cite{Dixit:2020ymh} 
however our scheme has significant potential to provide complementary and unique sensitivity at the high-frequency regime 
thanks to the insensitivity of the qubit excitation rate on the cavity volume.

The search scheme can be also directly benefited from 
the exponential advancement of the large-scale NISQ computers
led by, e.g., IBM~\cite{IBM_roadmap} or Google~\cite{google_roadmap}.
Since the requirements and the experimental setup are almost identical,
the improved qubit multiplicity and coherence in the NISQ machines 
will scale the typical sensitivity of this experiment as well.
Technically, it might be even possible to perform the
experiment within the existing NISQ machines in a parasitic manner 
by executing the circuits consisting only of readout.

Finally, we point out that the physics cases of the search 
can be widely extended beyond the hidden photon DM, such as the axion DM 
or other non-DM transient energy density such as dark radiation.

\vspace{2mm}
\noindent
\underline{\it Acknowledgments}: 
We thank to 
Atsushi Noguchi and Shotaro Shirai for 
discussing and providing the key feedback
on the experimental setup and the possible dark count contributions.
S.C.was supported by JSPS KAKENHI Grants No.\ 23K13093. T.I. was supported by JSPS KAKENHI Grants No.\ 23H04864 and JST PRESTO Grant No.\ JPMJPR2253, Japan. 
T.M. was supported by JSPS KAKENHI Grants No.\ 18K03608 and No.\ 22H01215. T.N.was supported by JSPS KAKENHI Grants No.\ 23K17688, No.\ 23H01182, and JST PRESTO Grant No.\ JPMJPR23F7, Japan.

\vspace{2mm}
\renewcommand{\theequation}{A.\arabic{equation}}
\setcounter{equation}{0}
\noindent 
\underline{\it Appendix}: In this
Appendix, we discuss the effects of cavity-like metallic package
surrounding the qubit.

The effect of the qubit packaging on the electric field requires careful consideration.
As it is often a metallic container electrically covering the qubit chip, 
it effectively becomes a microwave cavity.
The effective electric field $\vec{E}^{\rm (eff)}$ projected onto the cavity wall would vanish at the cavity wall,
since the secondary field ($\vec{E}^{\rm (EM)}$) is provoked
by the electrons in the metal reacting to the DM-included field $\vec{E}^{\rm (X)}$.
Importantly,
however, since the phases of $\vec{E}^{\rm (EM)}$ and $\vec{E}^{(X)}$ 
have different spatial evolution, the field cancellation
becomes imperfect off the wall. 
While $\vec{E}^{(X)}$ can be regarded as spatially homogeneous 
since the interaction of hidden photon is extremely weak 
and the effects of the cavity on the dynamics of hidden
photon can be safely ignored,
$\vec{E}^{\rm (EM)}$ is dependent on the position $\vec{x}=(x,y,z)$.  The field
configuration of $\vec{E}^{\rm (EM)}$ inside the cavity is obtained by
solving $\Box\vec{E}^{\rm (EM)}=0$ and $\vec{\nabla}\vec{E}^{\rm
  (EM)}=0$ simultaneously, with the boundary condition at the cavity
wall, $[\vec{E}^{\rm (EM)}_\parallel+\vec{E}^{(X)}_\parallel]_{\rm
  wall}=0$, where the subscript ``$\parallel$'' indicates vectors
projected onto the cavity wall.  

Consider, e.g., a case where 
a transmon is placed inside a cylinder-shaped cavity  
with its conductor plate perpendicular to the cylinder axis (defined as $z$-axis).
As only the $z$-component of the fields are relevant, we find
\begin{align}
  {E}^{\rm (EM)}_z (\vec{x})
  = - \frac{J_0 (m_X r)}{J_0 (m_X R)}
  {E}^{(X)}_z,
  \label{E(EM)}
\end{align}
where $r\equiv\sqrt{x^2+y^2}$ is the radial distance from the cylinder axis, $R$ is the radius of
the cylinder, and $J_0$ is the Bessel function of the first kind.
Note that $m_X$ is assumed not equal to any of the cavity mode frequencies here, 
which ensures $J_0 (m_X R)\neq 0$. 
Then, for the qubit located at $r=r_{\rm q}$, we obtain a $\kappa$ factor (defined in Eq.\ \eqref{kappa_def}):
\begin{align}
  \kappa^{\rm (cylinder)} = 1 - \frac{J_0 (m_X r_{\rm q})}{J_0 (m_X R)},
\end{align}
representing the enhancement or reduction on the total field due to the cavity effect.
Notably, when $m_X$ is close to one of the cavity mode frequencies $\omega_{\rm c}$ (which satisfies $J_0 (\omega_{\rm c} R)=0$),
$|\kappa|$ is enhanced in principle up about the $Q$-factor of the cavity mode.
Physically, this corresponds to the accumulation of the ``cancellation field" $\vec{E}^{\rm (EM)}$ in the cavity.
This causes a massive enhancement in the excitation rate particularly when a high-$Q$ cavity ($Q_{\rm cavity} \sim 10^{4-6}$) is considered since $\eta \propto \kappa \sim Q_{\rm cavity}$ (or $p_* \propto |\kappa|^2$ for $p_* \ll 1$).

While this cavity mode enhancement is of great interest for pursuing the ultimate sensitivity of the experiment, 
a careful analysis involving the detailed design of the cavity-qubit system is further required to conclude, which is left as a future work.
We just comment here
that, if the size of the cavity is of the order of $m_X^{-1}$,
$|\kappa|$ is expected to be $\sim O(1)$ or larger.

\bibliography{ref}

\end{document}